\documentstyle[preprint,aps,epsf]{revtex}                  
\input epsf.tex
\def\DESepsf(#1 width #2){\epsfxsize=#2 \epsfbox{#1}}
\global\let\epsfloaded=Y 
\begin{document}
\pagestyle{empty}                                      
\preprint{
\font\fortssbx=cmssbx10 scaled \magstep2
\hbox to \hsize{
\hfill $
\vtop{
 \hbox{ }}$
}
}
\draft
\vfill
\title{Dynamics of
Multi-Component, Multi-Field Quintessence}
\vfill
\author{Tzihong Chiueh}
\address{
\rm Department of Physics, National Taiwan University,
Taipei, Taiwan}

%
%
\vfill
\maketitle
\begin{abstract}

Self-consistent dynamics of the multi-component, $N$-field quintessence 
and gravity is formulated
as relativistic $N$-body dynamics, embedded in a static viscous flat space
and under the forces given by the interacting Lorentz scalar potential
via exchanges of {\it field} bosons.  
The Ratra-Peebles power-law potential of effective single-field
quintessence 
can be derived from this "microscopic" perspective.  In certain
situations,
the effective dynamics can be made identical to that of single complex
quintessence, except for that the overall $U(1)$ symmetry is
not manifestly broken.  The present
formulation
provides a convenient gauge for analyzing superhorizon 
perturbations and possibly for quantization of superhorizon fields and
gravity together.

\end{abstract}
%
%
\pacs{PACS numbers:
 }
%
%
\pagestyle{plain}
\section{Introduction}

The discovery of accelerating Hubble expansion has 
inspired intense investigations on various acceleration 
mechanisms\cite{1,2,3}.  
Conventionally, the cosmological constant of a suitable 
value does just fine to explain satisfactorily 
all existing data\cite{4,5,6}.  
However,
as the quality of cosmological observations are rapidly to be further 
improved in the near future, it is therefore timely to 
explore possibilities
beyond what has conventionally been taken for granted.  The 
proposal
of quintessence marks an important breakthrough in 
how one perceives the major constituents of the present-day
universe, 
which have conventionally been believed to be describable by
classical 
physics.  This new constituent is a field that permeates the entire 
space in a relative uniform manner, and is not a collection of 
matter particles.  Investigations on the quintessence dynamics
and its ultimate cosmological implications have been an active
research area in recent years\cite{7}.    

The field description in the context of cosmology
is not new, but was in the past limited to the physics of very
early universe,
where classical physics no longer holds.  The inflaton field 
that drives the early-universe
inflation is a pronounced example.  Reheating
is another example; it addresses how the radiation is created by
dissipating the rapid oscillation of the same inflaton field previously
driving the inflation\cite{8,9}.
In fact, the proposal of quintessence is well-motivated, by the recognition 
that the origin of the cosmological constant can be the ground state
of a certain scalar field.  If so, it is not too far-fetched to
consider the field to be now evolving on its way
to settle to the ground state on the cosmological time scale.

Somewhat different from the pure scalar field dynamics in the very early
universe, dynamics of quintessence must generally take into account other
evolving energy components, such as matter and radiation, in the universe.  
The presence of other energy components turns out to dilute the strengths 
of both the "drag" resulting from
Hubble expansion and the "force" given by the quintessence potential.
It is therefore of relevance to investigate how all coupled energy 
components, including gravity, evolve in time under a unified framework.
This paper first aims to address this issue. 

On the other hand, there may be more than one quintessence fields present in
the universe.  In general, these $N$ fields couple through some mutual 
interactions, in addition to the gravity.
The complexity of the multi-field problem rises drastically, due partly to 
the increasing number of $N$ degrees of freedom, and also to 
the $N^2$ mutual interactions.
Nevertheless, when the number of fields $N$ is sufficiently large 
the problem can become simple
again, since one may extract a few relevant dynamical variables
and derives an effective theory from this many-field problem.  In an analogy
to many-body problems, such an effective theory
is derived from a field theoretical approach.
If this program can be made successful, it will provide good
foundation for the effective potential 
of the effective quintessence
in terms of elemental mutual interactions
of individual quintessence fields. Furthermore, 
the mutual interactions should be regarded as exchanges of
some mediate bosons in the field space. 
The present paper mainly aims at
formulating a useful framework for such a many-field
problem.

It also turns out the present formulation, using $T(\equiv \ln(a))$ 
as the new time variable, is most convenient for analyzing cosmological 
perturbations in the long-wavelength limit\cite{10}.  In this
gauge, $T$ can remain unperturbed to the order of $k/H$, the ratio of 
Hubble radius to the wavelength of superhorizon perturbation, and
the cosmological perturbations can be conveniently 
perceived as the evolving difference
of two homogeneous and isotropic universes with slightly different
initial conditions.

In this paper, we do not wish to distinguish the quintessence
from the scalar fields that were active in the early universe 
before photons were produced (reheating), and regard these fields
al together as the quintessence.  That is, one may turn off the photon
fluid in the present formulation to address the inflaton dynamics.
The dynamics of single quintessence field
is formulated in Sec.(2).  Extension to multi-component, multi-fields
is presented in Sec.(3), where examples for deriving the
effective quintessence models are given.  We also touch upon
the cosmological perturbations with the present formulation
in Sec.(4).  Conclusions
and comments are given in Sec.(5).

\section{Single-Field Quintessence}

Assuming a homogeneous and isotropic universe, we have
the equation of motion for the scalar field $\phi$ under a 
potential $V(\phi)$ as
\begin{equation}
\ddot\phi+3H\dot\phi+{\partial V(\phi)\over\partial\phi}=0.
\end{equation}
The Einstein equations reads 
\begin{equation}
H^2\equiv({\dot a\over a})^2={2gV_{eff}\over 1-g\phi'^2}, \ \ \ 
{\ddot a\over a}=-2H^2+3g(2V_{eff}-(\rho_f+p_f)),
\end{equation}
where $g\equiv 4\pi G/3$, $\phi'\equiv d\phi/d T$ with $T\equiv\ln a$, 
$V_{eff}\equiv V(\phi)+\rho_f$, and $\rho_f$ and $p_f$ are the energy 
density and pressure of other fluid components, which may also 
include the space curvature.  The space curvature can generally be 
regarded as a fluid in this homogeneous and isotropic limit, with an
equation of state $p_f=-\rho_f/3$.  The time dependence of
$\rho_f$ is proportional to $a^{-2}$ and its value can be either
positive or negative depending on whether the space has negative
or positive curvature, respectively.

These coupled equations can be simplified if we let
the independent variable of Eq.(1) be $T$.  Thus, 
$\dot\phi=H\phi'$ and $\ddot\phi=H^2\phi''+\dot H\phi'$, with
$H$ and $\dot H$ replaced by Eq.(2).  Equations (1) and 
(2) combine to yield a single equation:
\begin{equation}
\phi''+(1-g\phi'^2)[3(1-{\rho_f+p_f\over 2V_{eff}})\phi'+
{1\over 2g}{\partial\ln V_{eff}\over\partial\phi}]=0.
\end{equation}
The peculiar feature of this equation is that $\sqrt{g}\phi'=\pm 1$ are
the fixed points, and solutions with $g\phi'^2-1<0$ or $>0$ cannot cross
the fixed points to change the signs.  The peculiar
behavior at the fixed points means that there exists a maximum
rate of coherent change of $\phi$ over the horizon.
This condition is
also reflected in the expression of $H^2$ in Eq.(2),
where $H^2$ may become singular.  If $V_{eff}$ does not change sign, 
the singularities
in $H^2$ can not be removed by having the numerator to vanish
simultaneously with the denominator.  
Such an unphysical result is automatically
avoided in Eq.(3), demanding that
$g\phi'^2-1$ must remain of the same sign.
As a non-negative $V_{eff}$ is of physical
relevance, we shall focus on the causal regime where
$g\phi'^2 < 1$ in the rest of this paper.

Having understood so, we may make an analogy of
the field dynamics to particle dynamics, in that one 
identifies $\sqrt{g}\phi'$ as the velocity $v$, 
$(1-g\phi'^2)^{-1/2}$ as the Lorentz factor $\gamma$, and
$u=\gamma\sqrt{g}\phi'$ as the momentum.  This nonlinear
transformation puts Eq.(3) into
\begin{equation}
{du\over dT}+3(1-{\rho_f+p_f\over 2V_{eff}})u
+{\gamma\over 2\sqrt{g}}{\partial\ln V_{eff}\over\partial\phi}=0.
\end{equation}
The first and third terms are the "01" components a second-rank
tensor and the second term is a vector.  It demonstrates that 
this dynamical system is not Lorentz covariant, because the
frictional force given by the Hubble expansion behaves as 
if the system contains a viscous medium which has already 
chosen a preferred reference frame.

Of particular interesting of Eq.(3) or (4) is the softening effect
to the field $\phi$ provided by another energy component $\rho_f$.
The field has a vacuum state $\phi_v$, where
$V(\phi_v)=V'(\phi_v)=0$.  The "force" 
$-\partial\ln V_{eff}/\partial\phi$ can be locally expanded as 
$-2(\phi-\phi_{min})[(\phi-\phi_{min})^2+(2\rho_f/V'')]^{-1}$.
It is analogous to a two-dimensional 
Coulomb attractive force given by a finite-sized 
charged distribution.
If the soft core does not exist, i.e., $\rho_f=0$, 
the "kinetic energy", 
or $\gamma$, will become infinity at the vacuum, so that
$H^2$ in Eq.(2) remains finite.  

To put the above into a quantitative perspective, we note that
$g^{-1/2}\sim E_{planck}$, the Planck energy.  Hence the
dimensionless "displacement" $\sqrt{g}\phi$ and velocity
$\sqrt{g}\phi'$ are all normalized to the Planck scale.  The
Planck scale thus typically implies a "relativistic" regime.  
However, the reverse statement is typically not true.
Energy scale much below the Planck regime can also
be "relativistic", as long as the dynamical time scale is sufficiently 
small so that $d/dT$ is large despite $\phi$ being small.  Such a
situation is commonly encountered when $V_{eff}\to 0$.

Despite not Lorentz covariant, the dynamics can still be put
in a Lagrangian framework that describes 
the self-consistent
evolution of field and gravity.  A straightforward
inspection shows that the
action for the coupled dynamics becomes
\begin{equation}
\int L dT = -\int e^{3T} dT\gamma^{-1} \sqrt{2g V_{eff}}
\end{equation}
By defining a new potential
$\Phi(\phi,T)\equiv\ln V_{eff}+6T$, we have here an action
of particle dynamics in the presence of a dilation field:
$\sqrt{2g}\int\exp(\Phi(\phi,T)/2) d\tau$, where $\Phi$ is a
dilaton field and $\tau$ is the proper time.  
In fact, the action given in Eq.(5) can be derived
rigorously by summing up the Lagrangians of quintessence
field, gravity and 
other energy components, as will be described in the next section.

Although there is no obvious conservation law for $\phi$ due
to cosmic expansion,
this Lagrangian nevertheless allows us to obtain a
Hamiltonian, which can yield an approximately conserved quantity.
This Hamiltonian $h$ is constructed by the usual procedure
\begin{equation}
h=\phi'(\partial L/\partial\phi')-L
=\gamma\sqrt{2gV_{eff}} e^{3T}.
\end{equation}
Actually, $h= H\exp(3T)$ according to Eq.(2).   

To show $h$ to be indeed useful, we obtain
the equation for $h$ by multiplying Eq.(5) by $u$ to get
\begin{equation}
{d\ln h\over dT}=
{3\over \gamma^2}(1-{\rho_f+p_f\over 2V_{eff}}),
\end{equation}
or after a straightforward algebra
\begin{equation}
{d\over dT}(V+u^2 V_{eff})=-6u^2 V_{eff}.
\end{equation}
It is clear that Eq.(7) is appropriate for the
"ultra-relativistic" regime, where the right-hand
side can be ignored and one obtains approximate conserved
quantities. 

The "ultra-relativistic speed" means that $\sqrt{g}\phi'\to \pm 1$,
implying $\sqrt{g}\phi\to \pm T$ in the large $|T|$ limit.
The "Lorentz" factor $\gamma$ can stays large as long as
$2gV_{eff} << h^2\exp(-6T)$.  
On the other hand, since $h=const.$ the Hubble parameter 
$H(=dT/dt)$ is proportional to $\exp(-3T)$ in this regime.
Solving for $T(t)(\equiv d\ln(a)/d\ln t)$, we find that
$\exp(T)\sim a(t)\propto t^{1/3}$, i.e., an universal expansion 
regardless of the form of $V(\phi)$ in this 
kinetic-energy-dominated regime.  This is the slowest 
asymptotic Hubble expansion rate one may hope to obtain.

From Eq.(4), one also sees that apart from the usual notion of
inflation where
a large range of flat $V(\phi)$ and a potential minimum are
needed for it to occur, the inflation can also
take place in a peculiar situation when $\rho_f=0$,
the potential has a maximum,
and the incoming kinetic energy is tuned to have
just the right value for it
to be completely consumed upon climbing to the potential top.
Since $v=0$ and $\partial\ln V/\partial\phi=0$, a state 
situated exactly at the separatrix also gives a constant $H$, 
thereby yielding inflation.  

Before leaving this simplest single-field dynamics, we 
shall also
look into the dynamics of typical relevant potentials.

\subsection{Exponetial potential}

The exponential potential, $V(\phi)=\exp(6\beta\sqrt{g}\phi)$, 
has interesting conformal symmetry, 
when $\rho_f= 0$ and $\beta=\pm 1$.  
The conformal transformation,
$\sqrt{g}\bar\phi=(1\pm v_t/1\mp v_t)^{1/2}\gamma_t(\sqrt{g}\phi-v_t T)
-T_0$ and $\bar T=(1\pm v_t/1\mp v_t)^{1/2}\gamma_t(T-v_t\sqrt{g}\phi)+T_0
-(1/6)\ln(1\pm v_t/1\mp v_t)$, transforms the original action
$\int dT\sqrt{1-g(d\phi/dT)^2}\exp[3(\pm\sqrt{g}\phi+T)]$ into 
$\int d\bar T
\sqrt{1-g(d\bar\phi/d\bar T)^2}\exp[3(\pm\sqrt{g}\bar\phi+\bar T)]$, 
where
the transform "velocity" $v_t$ and the shift $T_0$ are both constants.  
The transformed action assumes the same form as the original action,
and thus yields the same equation of motion.  
In fact, this particular case $|\beta|=1$
lies at the boundary of two distinctly different regimes, to be discussed
below.

When $|\beta|>1$, it is possible to make the above transformation
such that the $\bar T$ dependence vanishes 
by choosing $v_t=-\beta^{-1}$.  This is a
space-like regime where the Lagrangian becomes time-independent
and free of friction.  The conserved Hamiltonian $\bar h$ is a constant of
motion and becomes 
$\bar h\propto\bar\gamma\exp(3\sqrt{g}\bar\phi(\beta\pm 1))$. It allows
for a solution $\bar\phi=[1/3\sqrt{g}(\beta\pm 1)]
\ln[\cosh(3(\beta\pm 1)(\bar T-\bar T_0)/2)]$.

On the other hand, when $|\beta|<1$, one may choose $v_t=-\beta$ so
that the original exponential potential is transformed to become a 
$\phi$-independent vacuum energy, thereby yielding solutions that 
experience only the frictional force.  This is a time-like regime.
The Lagrangian becomes $\propto \bar\gamma^{-1}\exp(3(1\pm\beta)\bar T)$.
Since there is no applied force in the transformed coordinates, 
we have an asymptotic solution at rest, 
$\bar\phi=0$ and $d\bar\phi/d\bar T=0$.
It then follows that $H\propto\gamma\exp(3\beta\sqrt{g}\phi)\propto
\exp(3\beta v_t T)=\exp(-3\beta^2 T)$. Hence, 
the scaling factor
$a\propto t^{1/3\beta^2}$.  For $\beta^2<1/3$, these
solutions result in power-law inflation, for which the universe
undergoes accelerating expansion.

\subsection{Power-law potential}

The repulsive power-law potentials, $V(\phi)=V_0 \phi^{-n}$ for $n>0$,
can also yield inflation-like solutions, known as the 
Ratra-Peebles potentials\cite{11}.
The inertia, i.e., the second derivative of Eq.(3) or (4), can 
only be relevant only for a short while for the field to pick up 
the "speed".
But it will quickly become subdominant, and the field reaches
the terminal "velocity", where the frictional force balances the applied
force.  Moreover, as the terminal speed is always "non-relativistic",
Eq.(3) or (4) gives the solution, $\phi\propto T^{1/2}$ and hence
$H\propto \sqrt{V}\propto T^{-n/2}$, or $a\propto \exp(t^{2/(2+n)})$,
also an inflationary solution.
The above is for the case $\rho_f=0$.  When $\rho_f$ is finite and
even dominates $V_{eff}$, the balance of friction and applied force
then gives $V(\phi)\propto a^{-\kappa n/n+2}$, where we have let
$\rho_f\propto a^{-\kappa}$.  Apparently $V$ declines more slowly
than $\rho_f$ does, and hence the scalar field will eventually
dominate and the Hubble expansion approaches that of 
the fluid-free case.

The dynamics can be more complex when the power-law
potential is attractive, $V(\phi)\propto \phi^{2n}$, 
as it exhibits anharmonic damped
oscillations.  Nevertheless, the following general property
holds.  First, since the attractive force, 
$-\partial\ln V_{eff}/\partial\phi$,
becomes increasingly stronger in the regime where the average potential 
strength $\langle V\rangle>>\rho_f$, the oscillation frequency is 
increasingly higher.  When
the oscillation frequency is sufficiently high, one has a small
number, the ratio of the oscillation period to the Hubble time, as an
expansion parameter for a perturbative treatment in evaluating how
the oscillation amplitude declines in time.
(This high-frequency regime may also occur even when 
$\rho_f>>\langle V\rangle$ because $\rho_f$ can decrease rapidly in time.  
An accurate assessment of when the high-frequency regime holds will
be deferred to the end of this subsection.)  

To the leading order, the orbit in one oscillation can be viewed as
dissipationless, thereby allowing one to calculate the "ideal" orbit.
One may subsequently substitute the "ideal" orbit into the
damping term to evaluate how much damping it incurs to the orbit 
in one oscillation period.  The long-term damping
of the oscillation amplitude can therefore be evaluated to the
leading order after these two steps.

Ignore the right-hand side of Eq.(8) and one gets the momentum of the 
dissipationless oscillation orbit:
\begin{equation}
u^2={E(T)-V(\phi)\over V(\phi)+\rho_f(T)},
\end{equation}  
where $E$ is a quasi-static constant of motion, slowly
dissipated by the right-hand-side of Eq.(8).  
When $\rho_f\geq <V>$, the dynamics assumes 
non-relativistic motion and hence $E\approx \rho_f v^2+V(\phi)$.
That is, quintenssence acquires an effective mass of
$2\rho_f$.  Averaged over one oscillation period, 
Eq.(8) gives, to the leading order,
the rate of secular change for $E$:
\begin{equation}
{DE\over DT}\approx -6{\int dT (V+\rho_f)u^2\over \int dT}
=-6{\int d\phi \sqrt{E-V(\phi)}\over
\int d\phi(\sqrt{E-V(\phi)})^{-1}}=
-3{dE\over d\ln(\int d\phi\sqrt{E-V(\phi)})}.     
\end{equation}
Here we have used the non-relativistic approximation 
$dT=\sqrt{g}d\phi v^{-1}$, $D/DT$ is the derivative
on the slow time scale and the second equality is merely an algebraic 
equality.  Since the right-hand side of the 
third equality is a derivative on the slow quantity,
Eq.(10) can thus be trivially integrated to yield 
\begin{equation}
T(E)-T_0=-{1\over 3}\ln(\int d\phi\sqrt{E-V(\phi)}),
\end{equation}
where $T_0$ is also an integration constant.  

We now apply the above results to the attractive power-law potential,
$V(\phi)\propto \phi^{2n}$. The integral
$\int d\phi\sqrt{E-V(\phi)}\propto E^{(n+1)/2n}$, and 
$T(V_0)=-\ln E[(n+1)/6n]+const.$, or
$E\propto a^{-6(n+1)/n}$.  That is, a quadratic potential, 
$n=1$, yields $E\propto a^{-3}$, similar to a matter fluid 
and a quartic
potential, $n=2$, gives $E\propto a^{-4}$, similar 
to a photon fluid.

Having Eq.(11), we may further evaluate 
$Dt=\int dt (=\int dT H^{-1})$, so that the averaged 
Hubble expansion rate $\bar H=DT/Dt$ has the expected expression:
\begin{equation}
\bar H={DT\over Dt}=\sqrt{2g(E(T)+\rho(T))}.
\end{equation}
Equations (11) and (12) together permit the
deceleration parameter to also be evaluated:
\begin{equation}
q(T)\equiv{1\over 2a\bar H^2}{D^2a\over Dt^2}
={1\over 2}{D\ln(E(T)+\rho_f(T))\over DT}+1.
\end{equation}

All the above results for the long-term evolution is based on 
the assumption that oscillation frequency is much larger
than the Hubble expansion rate.  
It is evident from Eqs.(3)
or (4) that when $\rho_f$ is much greater than $V(\phi)$,
the force strength is substantially reduced by the factor 
$V(\phi)/\rho_f$, thus reducing the oscillation frequency
by $\sqrt{V(\phi)/\rho_f}$.  Hence, there is a limit,
beyond which the above results fail to hold.  The regime in 
which
this may happen is always "non-relativistic", and one may 
approximate the right-hand side of Eq.(4) by $-\omega^2\phi$,
where $\omega$ is the dimensionless oscillation frequency,
to estimate $\omega\sim
[(dV/d\phi)/2\sqrt{g}\phi(\rho_f+\phi)]^{1/2}$.
Again, using the power-law potential, $V\propto\phi^{2n}$, we
find that $\omega\propto \phi^{-1}\propto a^{3/(n+1)}$ if 
$\langle V\rangle>>\rho$, and that 
$\omega\propto\phi^{n-1} a^{3(1+w)/2}\propto a^b$
when $\rho >> \langle V\rangle$, where 
$b\equiv 3(1+w)/2-3(n-1)/(n+1)$ and the fluid equation 
of state $w=p_f/\rho_f$ has been assumed.  In the former
case, all confining ($n>0$) potentials give the desired 
high frequency.  
In the latter situation, we, however, need $b\geq 0$, or
$n\leq (3+w)/(1-w)$.  For photon fluids, one needs $n\leq 5$,
and for matter fluids, $n\leq 3$. 

\section{Dynamics of multi-component N fields}

\subsection{General framework}

The success in describing the field dynamics by relativistic particle 
dynamics motivates us to go one step further, and 
extend this formulation to the dynamics of 
multi-component, multi-fields.  Here, the component of a field refers
to the requirement that it has a geometric structure as a vector.  

As an illustration for how
this can be done, we shall first consider a simple example of
multi-component, single-field quintessence with no fluid component
present.  In this case, 
the potential $V$ is a scalar and cannot be an arbitrary 
function of $\phi_i$, where $i$ is the component index.
A particularly interesting potential
is $V({\bf\phi})=V_0\exp[{\bf\beta}\cdot{\bf\phi}]$, where ${\bf\beta}$ 
satisfies ${\bf\beta}^2=1$.  In this particular case, 
the constant vector ${\bf\beta}$ picks out a particular direction
of ${\bf\phi}$.  Along this direction, one may perform the 
conformal Lorentz$+$shift
transformation discussed in Sec.(2) for the single-component
quintessence,
and obtain an invariant action under such a transformation.   The
action is also invariant to the conformal helical (rotation$+$shift)
transformation along the ${\bf\beta}$ direction. 

It turns out that the action of this particular $N$-component, 
single-field 
quintessence can also be the action of the $N-M$ component,
$M$-field quintessence.
This is due to that the Lorentz factor (c.f., Eq.(5)) in the action
contains the "velocities" of all fields, as will be elaborated below. 
Hence, the above conformal symmetry still holds for the
$N-M$ component, $M$-field quintessence, as long as
$\sum_i{\bf\phi}_i^2=1$.  

The above example shows that
when the potential $V$ assumes some particular forms, 
there can be
no distinction between the multi-component single-field 
and the multi-field.  To have a distinct
difference, the potential should have a more general form.  Below, we
consider such a situation.  For the purpose of illustration,
we are confined to the two-component 
fields, which can be conveniently
described by complex fields.  Fields with more than two 
components can be easily generalized.

Let $\psi_i$ be the $i$-th of a system of $N$ two-component fields.
We are interested in an autonomous system of interacting 
$N$ fields, where the interacting potential can be written as 
$V_{i,j}=V(|\psi_i-\psi_j|)$ with $i\neq j$.
The action for the system of $N$ fields, the gravitation 
and a fluid component can thus be written as
\begin{equation}
S_0=-l^3\int a^3(t)dt [(\sum_i{\dot\psi_i^2\over 2}-\sum_{i,j} V_{i,j})+
({R\over 12g})+(\lambda({n_f^2\dot\theta_f^2\over 2(\rho_f+p_f)}-
{\rho_f+p_f\over 2})+p_f)],
\end{equation}
where $l^3$ is the space volume, $R(\equiv(1/2)[(\dot a/a)^2+(\ddot a/a)])$ the 
scalar curvature, $\lambda$ the Lagrangian multiplier, and $n_f$ 
and $\theta_f$ are the number density and
velocity potential of the fluid, respectively.  

The Lagrangian multiplier
$\lambda$ is introduced to defined what the fluid $4$-velocity $U_\mu$ is
meant to be.
In fact a full fluid Lagrangian has been explored before\cite{12}, 
and here we consider 
only the spatially-uniform limit.  Variation of $S_0$ with respect to 
$\lambda$ gives the constraint: 
\begin{equation}
({n_f\dot\theta_f\over \rho_f+p_f})^2=1;
\end{equation}
variation with respect to $\theta_f$ yields 
a conservation law,
\begin{equation}
{a^3n_f^2\dot\theta_f\over \rho_f+p_f}=s^2,
\end{equation}
where $s$ is a real constant.  Since the $4$-velocity satisfies
$U_\mu U^\mu=1$, the constraint, Eq.(15), defines the fluid 
$4$-velocity as
$U_\mu=n_f\partial_\mu\theta_f/(\rho_f+p_f)$.   
It then follows that Eq.(16) gives
$a^3 n_f U_0=s^2$; as the fluid is rest in the comoving-frame of
cosmic expansion,
this becomes $a^3 n_f=s^2$, i.e., the conservation of particle
number density in an uniformly expanding fluid.  
Variation with respect to $n_f$ gives 
$\partial(\lambda\rho_f+(\lambda-1)p_f)/\partial\ln n_f
=\lambda(\rho_f+p_f)$.  Note that the ideal-gas law demands
$\partial\rho_f/\partial\ln n_f=\rho_f+p_f$, therefore fixing 
$\lambda=1$.

The Einstein equations is derived from
this action by variation with respect to the scale factor $a$, 
and it gives, as before,
\begin{equation}
H^2=g{2\sum_{i,j}V_{i,j}+\rho_f-p_f\over 1-g\sum_i(\psi'_i)^2-
g[n_f^2(\theta_f')^2/\rho_f+p_f]}, \ \ 
{\ddot a\over a}=-2H^2+3g(2\sum_{i,j}V_{i,j}+\rho_f-p_f),
\end{equation}
where $\psi_i'\equiv H^{-1}\dot\psi_i$ and  
$\theta_f'\equiv H^{-1}\dot\theta_f$.

Now, replacing $dt$ by $dT/H$, substituting $H$ of
Eq.(17) into $d/dT$ and recognizing $R=6g(\rho_{total}-3p_{total})$,
we finally obtain the effective $N$-field and fluid action
for Eq.(14), much like how Eq.(5) is obtained.  In this 
effective action the gravity is manifestly hidden:
\begin{equation}
S_0=-l^3\int e^{3T} dT
(\bar\Gamma)^{-1}\sqrt{g[2\sum_{i,j}V_{i,j}+\rho_f-p_f]},
\end{equation}
where $(\bar\Gamma)^{-1}\equiv \sqrt{1-g\sum_i(\psi'_i)^2-
g n_f^2(\theta_f')^2/(\rho_f+p_f)}$.

The action recovers Eqs.(15) and (16), and again the appropriate 
ideal-gas law is needed to fix $\lambda=1$.  The non-trivial
extension of this action from that of Eq.(5) is the 
appearance of an effective Lorentz factor $\bar\Gamma$,
which is {\it not} the sum of individual Lorentz factors $\gamma_i$,
as one may have expected.  Rather, it is the sum of "kinetic energy"
of all fields that contributes to $\bar\Gamma$.
When $N$ is sufficiently large, it is obvious that 
$g(\psi_i')^2 << 1$ in order for $\bar\Gamma$
to be real, thus demanding each individual field to be 
"non-relativistic".  In addition, it is straightforward to derive
the corresponding Hamiltonian $h$, as we did for Eq.(6), 
and the relation $h=H\exp(3T)$ still holds.

Although Eq.(18) contains a complete Lagrangian for describing the 
coupled dynamics of fluid and quintessence in terms of the logarithmic
scaling factor $T$, the fluid and quintessence in fact form a master-slave
system.  That is, in terms of the variable $T$, the fluid dynamics
is independent of the presence of quintessence, but the quintessence
dynamics is governed by the presence of fluid.  As the fluid dynamics can be
easily solved when the equation of state is given, it is therefore 
possible to simplify the algebra by focusing on the quintessence piece
of Eq.(18), with the fluid variables $\rho_f$ and
$p_f$ treated as known functions of $T$.  This is the 
strategy that was adopted in Sec.(2) and it gives a much simpler 
and practical result:
\begin{equation}
S_q=-l^3\int e^{3T} dT
\Gamma^{-1}\sqrt{2g V_{eff}},
\end{equation}
with the fluid satisfying
\begin{equation}
{d\rho_f\over dT}=-3(\rho_f+p_f),
\end{equation}
where $V_{eff}=\sum_{i,j}V_{i,j}+\rho_f(T)$ and
$\Gamma^{-1}\equiv \sqrt{1-g\sum_i(\psi'_i)^2}$.
Moreover, the Einstein equations, Eq.(17), now become
Eq.(2), with the "Lorentz" factor in Eq.(2) replaced by
the multi-field $\Gamma$ defined here.

We would like to mention in passing that the way by which 
the multi-field version of Eq.(3) is derived from
this multi-field quintessence action, Eq.(19), has 
a slight twist.  Extend Eq.(3), $G=0$, to the expected 
multi-field 
version, $G_i=0$.  The Euler-Langrange equation of
Eq.(19) in fact yields 
$\sum_j[(\delta_{ij}-P_{ij})+\Gamma^{-2}P_{ij}]G_j=0$, where 
$P_{ij}\equiv\delta_{ij}-\psi_i'\psi_j'/\sum_k(\psi_k')^2$,
a projection operator perpendicular to $\psi_i'$.  
Since $(\delta_{ij}-P_{ij})+\Gamma^{-2}P_{ij}$ 
is non-vanishing, this equation 
simply implies that the component of $G_i$ parallel
to $\psi_i'$ and those perpendicular to $\psi_i'$
both vanish, and therefore $G_i=0$.  The projection
into two components relative to the "velocity" manifestly
indicates that the special-relativistic effect is in action for 
producing anisotropic forces.
In particular, Eq.(4)
should be rewritten from $du/dT=F$ to
$du_i/dT=\sum_j[(\delta_{ij}-P_{ij})+\Gamma^{-2}P_{ij}]F_j$
in the multi-field case.
This feature of dynamics given by
Eq.(19) has been discussed extensively in the past, and
the potential $V$ has been known as the Lorentz scalar 
potential\cite{13}.

\subsection{Coulomb field gas}

One may be contented with this Lagrangian, where $V_{i,j}$ is a 
given function.  But one may also
be somewhat more ambitious and asks whether there may be a 
fundamental dynamical equation that accounts for $V_{i,j}$.  
Actually, such a question is well-motivated,
since these $N$-fields are like $N$-particles.  
It is natural for $N$-particles
to interact via exchange of some mediate bosons, 
such as the familiar 
photons in electromagnetic interactions, and these mediate 
bosons obey their own dynamical equations.  Here, we shall 
regard $V_{i,j}$ as the manifestation of such {\it scalar} 
mediate bosons of zero frequency.  
To attack this problem, we will need to go beyond the original discrete
field $\psi_i$ and consider $V$ to be the field generated by and interacting with
the discrete
$\psi_i$, which in the continuum limit can be replaced by  
$\int d^2\psi \delta(\psi-\psi_i)\psi$ where
the Dirac $\delta$-function is in use.  The simplest version for
$V(\psi)$ dynamics has an action:
\begin{equation}
S_V={l^3 N^2\over 4\pi e^4}\int d^2\psi\int e^{3T} (dT/H) 
[\kappa^{-2}({\partial V\over \partial T})^2
-({\partial V\over\partial \psi})^2],
\end{equation}
where $e^4$ is the coupling constant with $e$ 
having a dimension of energy,
and $\kappa$ is also a constant having a dimension of energy as well.

A good working hypothesis for the constant $\kappa$, which relates 
the metrics of $\psi$ to the expansion factor $T$, is that 
$\kappa\sim g^{-1/2}$, thus governed by the physics of gravity.  
On the other hand, the field-field coupling constant $e^4$ is 
likely governed by physics other than gravity, and we shall
leave its magnitude as a free parameter.

Moreover, one needs to extend the action $S_0$ of Eq.(14) to a
more general one:
\begin{eqnarray}
S&=&l^3\int e^{3T}(dT/H) 
[\int d^2\psi \sum_i \delta(\psi-\psi_i)
(H^2{(\psi_i')^2\over 2}-
\int d^2\psi'\sum_j \delta(\psi'-\psi_j)V(\psi,\psi') )\nonumber\\ \
&+&{R\over 12 g}+ {\lambda\over 2}({n_f^2\dot\theta_f^2\over
\rho_f+p_f}-(p_f+\rho_f))+p_f]+S_V,
\end{eqnarray}
Variation of $S$ with respect to $V$ gives the
equation of motion for $V$:
\begin{equation}
{e^{-3T}H\over\kappa^2}[{\partial\over\partial T}({e^{3T}\over H}) 
{\partial V\over\partial T}]
-{\partial^2 V\over\partial \psi^2}=2\pi {e^4\over N}
\sum_j \delta(\psi-\psi_j).
\end{equation}
Equation (23) shows that
there exists an retarded effect in $V$, and hence 
the functional form $V(|\psi_i-\psi_j|)$, suitable only for
static fields, is over-simplified.  However, as has been 
mentioned earlier, each individual
field generally has a "non-relativistic" velocity $\sqrt{g}\psi_i'$ 
in the large-$N$ limit and moreover $\kappa^2\sim g^{-1}$.  
Therefore, to the leading order of 
$\sqrt{g}\psi_i'$, the field $V$ is approximately quasi-static.
This is reminiscent of the situation where a collection of slowly moving
particles of same charges generate mostly the electrostatic 
fluctuations and only little electromagnetic fluctuations.   
Ignoring the time derivative of $V$ in Eq.(23),
we obtain a static, axi-symmetric solution for 
the potential contributed by the discrete source $\psi_i$:
\begin{equation}
V(|\psi-\psi_i|)=-{e^4\over N}\ln|\psi-\psi_i|,
\end{equation}
or the Green's function for the two-dimensional Poisson equation.  

Finally, we sum up all terms in $S$ 
explicitly to obtain the total effective action.
It follows after an integration by part over $\psi$ that 
the total action is almost identical
to $S_0$ of Eq.(18), except for replacing $V_{i,j}$ by
$(1/2)V_{i,j}$.  This
factor $1/2$ is characteristic of a self-consistent electric 
or gravitational potential built up from a vacuum.
The contribution of interaction potential
to the Einstein equations is also given by the new potential,
$(1/2)V_{i,j}$, rather than the original one.  The above shows
that quintessence, its interactions, gravity and fluid can all be
incorporated in a single action.

But, as has been mentioned earlier, the full action with the fluid
included has no obvious advantage for algebraic manipulations, and
hence we shall from now on remove the fluid action from the quintessence
action.  Equation (20) will be needed for providing $\rho(T)$ that 
appears in the quintessence action.  Repeating
the same procedure, we finally arrive at an explicit form for the
quintessence effective action, Eq.(19):
\begin{equation}
S_q=-\int e^{3T} dT\Gamma^{-1}
\sqrt{2g[{e^4\over 2N}\sum_{i,j}\ln(|\psi_i-\psi_j|)+\rho_f]},
\end{equation}
from which the equation of motion for each individual field 
$\psi_i$ can be derived.  

The above exercise demonstrates that the quintessence potential $V$ 
can be given a more fundamental origin, and
in this particular example we obtain a system of $N$ fields interacting 
among themselves like a two-dimensional Coulomb gas.  
If the fields have three 
components, the interacting system will be like a 
three-dimensional Coulomb gas.  Generally, the 
$m$-component fields will behave like a $m$-dimensional Coulomb gas.
The Coulomb gas self-repels and can maintain 
a fairly uniform density.  It can therefore be
a relatively simple N-body dynamical system for 
the description of the cosmic evolution,
if the Coulomb gas is initially cold and uniformly distributed.  
For a finite $N$, the uniformly (in $\psi$ coordinate) distributed 
fields occupy within a uniform disk of finite area $\pi q^2$.  
The disk experiences a uniform radial force, except at the edge. 
The outward force at the edge,
$-\partial V/\partial q\equiv -|\partial V/\partial\psi|_{|\psi|=q}$, is
proportional to $q^{-1}$, characteristic of a 
two-dimensional Coulomb force.  The radius $q$ thus becomes 
the only relevant dynamical variable for
describing the evolution of the field energy density, 
and we recover the single-field dynamics.

Specifically, Eq.(25) contains $\Gamma^{-1}=
\sqrt{1-g\sum_i(\psi_i')^2}$ and $\sum_{i,j}V_{i,j}$.
In a uniform cold disk, the velocity is proportional to the radius
$|\psi|$, much like the Hubble expansion, and hence 
$\sum_{i}(\psi'_i)^2=(d\ln q/dT)^2(2\pi\int_0^q |\psi|^3 d|\psi|)
(N/\pi q^2)=(N/2)(q')^2$.  On the other hand, the
force at the edge can be obtained easily by using the 
Stokes theorem for a disk of uniform surface 
density $e^4/\pi q^2$:
that is, $V(q)=e^4\ln(1/ q)+c_0$, where $c_0$ is a constant
serving to offset the ground-state energy.   
The effective action
for the single-field description of this $N$-field problem now
becomes
\begin{equation}
S_{q,eff}=-\int e^{3T}dT (1-{Ng(q')^2\over 2})^{1/2}
[2g(e^4\ln({1\over q})+c_0+\rho_f)]^{1/2},
\end{equation}
or upon being properly normalized, 
\begin{equation}
S_{q,eff}=-\int e^{3T}dT (1-g(\bar q')^2)^{1/2}
[2g(e^4\ln({1\over\bar q})+\rho_f)]^{1/2},
\end{equation}
where $\bar q\equiv q\sqrt{N/2}$ and the constant $c_0$ is so
chosen that the maximum $\bar q$ is unity, a requirement that the
initial condition must satisfy.
This action describes the $m=2$ case of the Ratra-Peebles 
Potentials\cite{11}, $V(q)\propto q^{-(m-2)}$, for the
single-field quintessence, where late inflation can always occur.  
In fact, all Ratra-Peebles potentials of 
integer $m-2$ can be generated by
this $N$-field model of $m$-component fields.

\subsection{Warm Coulomb field gas}

The above cold-gas model is of the simplest type. 
A reasonable extension of it
can be for an initial condition where all fields are still uniformly
distributed but having thermal velocities.  In this case,
we need to consider a mean-field phase-space distribution 
$f(\psi,\psi')$ to describe the
$N$-field kinematics when $N$ is large.  
That is, while $V(\bar q)$ remains to be $-e^4\ln(\bar q)$ as is for the
cold gas, the quantity $\sum_i(\psi_i')^2$ demands a more elaborateed
treatment.  In the large $N$ limit, the long-range nature of
Coulomb interactions allows each field
to have a large mean-free path and "sees" only the mean field $V_{mf}$,
which equals to $V(\bar q)$ at $q=\bar q$. 
According to Eq.(3), the distribution function therefore satisfies
\begin{equation}
{\partial f\over\partial T}+\psi'\cdot{\partial f\over\partial\psi}-
\Gamma^{-2}[{1\over 2g}{\partial\ln V_{mf}\over\partial\psi}
-3(1-{\rho_f+p_f\over V_{eff}})\psi']\cdot
{\partial f\over\partial\psi'}=0.
\end{equation} 

We follow the mean-field relativistic kinetic theory:
\begin{equation}
{1\over N}\sum_i=\int d^m\psi\int d^m\psi' f(\psi,\psi')
=\int d^m\psi n_b, 
\end{equation}
where $m$ is the number of field components and 
the number density $n_b$ has been normalized to unity;
\begin{equation}
{1\over N}\sum_i\sqrt{g}\psi_i'
=\int d^m\psi[\int d^m\psi'\sqrt{g}\psi' 
f(\psi,\psi')]=\int d^m\psi [n_b v_b], 
\end{equation}
where $v_b$ is the local bulk velocity; 
\begin{equation}
{1\over N}\sum_i g(\psi_i')^2=\int d^m\psi\int d^m\psi'
[g(\psi')^2 f(\psi,\psi')]=
\int d^m\psi [n_b v_b^2+2{U_\psi\over N}],
\end{equation}
where $U_\psi$ is the thermal energy.
Here, we have
assumed that $f$ has an isotropic velocity distribution.  

The appearance of the thermal energy $U_\psi$ adds an extra 
dynamical variable to this problem, and it needs to be solved
self-consistently.  This task can be relatively easy to handle
when $N$ is large.  To a good approximation, 
each field freely streams, much like an ideal-gas
particle, so that the adiabatic ideal-gas law provides
the needed description for $U_\psi$.  For uniformly
distributed $N$-fields, we may use a scaling factor
$b(t)$ to account for the bulk motion, much like
how the scaling factor of the Hubble flow $a(t)$.  That is,
$\psi_i=b(t) x_i$ and $\dot\psi_i=H\psi'=\dot b x_i+b\dot x_i
=H(v_i+\delta v_i)/\sqrt{g}$, where $\delta v_i$ is the
"thermal" velocity.  
In fact, the individual field $\psi_i$ satisfies the 
original field equation,
\begin{equation}
0=\ddot\psi_i+3H\dot\psi_i+{\partial V_{mf}\over\partial \psi_i}
=[\ddot b+3H\dot b+(V_{mf}''/2) b]x_i
+[b\ddot x_i+\dot x_i(2\dot b+3Hb)].
\end{equation}
The first bracket vanishes, as it describes the bulk
flow or the uniform expansion.  The second bracket
must therefore also vanish, thereby allowing $\dot x_i$
to be solved: $\dot x_i\propto b^{-2} a^{-3}$.  Furthermore,
the dynamical variable, the disk radius $q$, is proportional
to the field scaling factor $b$, and it thus follows that the thermal
energy 
\begin{equation}
U_\psi=({1\over 2})\sum_i \delta v_i^2=\alpha^2 q^{-2} a^{-6},
\end{equation}
where $\alpha$ is a real constant.  

Since 
$\sum_i\delta v_i^2$ has already been solved, there is no need
to consider the "thermal" component of $g(\psi'_i)^2$
appearing in $H^2$ of Eq.(17), and thereafter in $\Gamma$
of the action $S$ in Eqs.(19) and (25).  This procedure is identical to
how we treated the fluid component in converting $\bar\Gamma$
of Eq.(18) to $\Gamma$ of Eq.(19).  
Substituting $g\sum_i(\psi_i')^2=\sum_i v_i^2+H^{-2}U_\psi$ 
into Eqs.(17),
we have the new expressions for the Einstein equations:
\begin{equation}
H^2={2g \bar V_{eff}\over 1-Ng(q')^2/2}, \ \ \ 
{\ddot a\over a}=-2H^2+3g[2\bar V_{eff}-(\rho_f+p_f)],
\end{equation}
where $\bar V_{eff}\equiv V_{mf}(q)+\rho_f+U_\psi(q,a)$, 
$\sum_i v_i^2=Ng(q')^2/2$.  We also have
a new expression for the effective quintessence action:
\begin{equation}
S_{q,eff}=l^3\int e^{3T} dT\sqrt{2g\bar V_{eff}(1-Ng(q')^2/2)},
\end{equation}
Thus, very different from what Eq.(19) or (25) may have suggested,
the Lorentz factor here contains only the bulk
flow, and the "thermal" energy contribution is absorbed
into $V_{eff}$.  The functional form of $V$ depends on
what kind of disk configurations under consideration.  
The new effect arising from
$U_\psi$ will lead to a qualitatively very different result
when the potential is attractive, where the nearly balance of the 
repulsive
force of $U_\psi$ and the attractive force of $V$ gives arise to
quasi-equilibrium trajectories, which on one hand oscillate on
the fast time scale and on the other hand sink to the
potential bottom on a secular time scale.

It is of little surprise to find that the effective action of
Eq.(35) also describes the dynamics of a single
complex field \cite{14}, the so-called "spintessence"\cite{15}, 
where we let $N=1$ and $m=2$.  The new effect pertaining to
$U_\psi(q,a)$ is identical to the "centrifugal potential" of
"rotational motion" of the single complex quintessence.  
This is because the "thermal motion"
allows the trajectory of each individual
field to carry a finite "angular momentum" relative to 
other fields, and these transverse "velocities"
give rise to the "centrifugal forces".  
However, unlike "spintessence", 
where the $U(1)$ symmetry is globally destroyed to a large
degree, the "thermal motion" of $N$ fields break
the $U(1)$ symmetry to a much lesser degree.  In particular, when
the coarse-grained average of fields of a large $N$ is taken,
the $U(1)$ symmetry is nearly restored, at worst broken
by $N^{-1/2}$ due to the Poisson statistics.

\subsection{Hamilton-Jacobi Equation}

We proceed to discuss how the present formulation can be extended
to the Hamilton-Jacobi theory, which will prove to be useful for 
considerations of the cosmological perturbations of superhorizon size
to be discussed later.  We shall now return to the general expression, 
Eq.(18), for the many-field action.
Given the Lagrangian in Eq.(18), we find the canonical momenta of the fields and
fluid to be
\begin{equation}
\pi_i=g\bar\Gamma\psi_i'\sqrt{2V_{eff}-\rho_f-p_f}e^{3T}, \ \ \ 
\pi_f=g\bar\Gamma({n_f^2\theta_f'\over \rho_f+p_f})
\sqrt{2V_{eff}-\rho_f-p_f)}e^{3T},
\end{equation}
where the Lagrangian multipier $\lambda$ has been set to its due value,
$1$.  As the Hamiltonian $h=H e^{3T}$, it follows from Eq.(17) that
\begin{equation}
h^2=\sum_i\pi_i^2+({\rho_f+p_f\over n_f^2})\pi_f^2+
g(2V_{eff}-\rho_f-p_f)e^{6T}.
\end{equation}
We now employ the Hamilton-Jacobi theory to rewrite
this Hamiltonian into
\begin{equation}
g^2({\partial S\over\partial T})^2=\sum_i({\partial S\over\partial\psi_i})^2
+({\rho_f+p_f\over n_f^2})({\partial S\over\partial\theta_f})^2
-g(2V_{eff}-\rho_f-p_f)e^{6T},
\end{equation}
where the canonical momenta $\pi_i$ and $\pi_f$ 
are cast into $\partial S/\partial\psi_i$ and 
$\partial S/\partial\theta_f$ respectively, 
the Hamiltonian $h\equiv g\partial S/\partial T$ and $S$ is
the Hamilton principal function.  The solution to this 
Hamilton-Jacobi equation has the form $S=S_q(\psi_i,T)+
S_f(\theta_f)$, where $S_q=W_q(\psi_i,T)\exp(3T)$.  The fluid
part $S_f$ on the right-hand side of Eq.(38) cancels by itself
since the time component of the
fluid $4$-velocity, $n_f\dot\theta_f/(\rho_f+p_f)$,
equals to unity, and it yields $\pi_f=\sqrt{g}n_f\exp(3T)$. 
Furthermore, the fluid momentum $\pi_f$ turns out to be 
a constant since the fluid solution obeys 
$n_f\propto \exp(-3T)$, thus allowing $S_f$ 
consistently to be independent of $T$ explicitly.
Equation (38) then becomes
\begin{equation}
\sum_i({\partial W_q\over\partial\psi_i})^2- 
g^2({\partial W_q\over\partial T}+3W_q)^2=-2g (V+\rho_f(T)).
\end{equation}

At any given time slice $T=T_n$, the phase function $S_q$ is
a function of $\psi_i$, and $\partial W_q/\partial\psi_i$ is parallel
to the velocity $\psi_i'$.
Any virtual displacement 
$\Delta\psi_i\equiv \psi_i^{(\alpha)}-\psi_i^{(\beta)}$ lying on the
$Nm-1$ dimensional hypersurface $W_q(\psi_i,T_n)=const.$ is
perpendicular to the local velocity $\psi_i'$, where $\psi_i^{(\alpha)}$
and $\psi_i^{(\beta)}$ are two adjacent instantaneous trajectories
at $T=T_0$.  At the next instant, $T=T_0+\delta T$, the hypersurface
evolves and so do the two trajectories $\psi_i^{(\alpha)}$ and 
$\psi_i^{(\beta)}$.  However,
at this next instant the two trajectories may not land on the same 
constant-$W_q$ hypersurface.  In order for the two instantaneous 
orbits be on the same hypersurface, they must be evaluated at different
times.  The progressively asynchronous orbits on the constant-$W_q$
hypersurfaces occur regardless whether the fluid is present.
  
The notion of the constant-$W_q$ hypersurface is useful only 
when $\Delta\psi_i$ is sufficiently small, i.e.,
in the regime of linear perturbations.  The reason is that as the multi-field
dynamics is likely chaotic, $W_q$ is generally not
an analytical function of $\psi_i$.
Generally speaking, the constant-$W_q$ surface should consist of 
foliated patches jointed by kinks and cusps, and
an analytical $W_q$ exists only in some immediate neighborhoods of
a given location $\psi_i$.  Hence, as long as $\Delta\psi_i$ is small, 
two trajectories can always stay in the same foliation patches at 
different times.  In the next section, we will make use of this notion to
discuss the cosmological linear perturbations in the comoving gauge.

\section{Cosmological Perturbations}

The above formulation using $T(\equiv \ln a)$ as the time variable
is a natural choice for describing the coupled dynamics 
of the background scalar field and gravity.  
It turns out that such a choice
is also convenient for analyzing the cosmological perturbations
of space curvature\cite{10}.

In the very long-wavelength
limit, for which the wavelength $k^{-1}$ is much greater than the 
Hubble radius $H^{-1}$,
the quintenssence and matter perturbations, $\delta\psi_i$ and
$\delta\rho_f$,
can be regarded as spatially homogeneous.
The energy-momentum tensor of each matter 
component is diagonal in the limit $k/H\to 0$, 
since the off-diagonal component involves 
two spatial derivatives and hence is of order $(k/H)^2$ and
the space-time component is of order $(k/H)$.  
Such a spatially homogeneous diagonal energy-momentum 
tensor leads to diagonal metrics.  
Thus, superposed with super-horizon
perturbations of $k/H\to 0$, 
both scalar field and fluid satisfy the same equations of 
motion as those of the background derived in the last section.
The cosmological perturbations are therefore the deviation of different
trajectories of slightly different initial conditions.

In fact, one can do better than the above zero-th order (of $k/H$)
considerations and retains
the perturbed quantities of order $k/H$. 
It is well known that the metric fluctuations have gauge degrees
of freedom, and some gauge choices may be more convenient than others.
As has been shown by Sasaki and Tanaka\cite{10}, the gauge that keeps 
the time variable $T$ unperturbed turns out 
to be the convenient gauge in allowing one to calculate perturbations
of order $k/H$, specifically for the perturbed space curvature $R$ and 
the traceless stress.  
In this gauge,
the perturbed shift (space-time component) of the metrics 
vanishes 
and the perturbed space compression (trace of space component) of 
metrics remains a constant.
The equation of motion derived above, describing
the "00" component and the trace of the space components of 
the energy-momentum tensor, remains
valid even up to $O(k/H)$ and is decoupled from other tensor components
of size $(k/H)$, namely, the perturbed space curvature and traceless stress.
The great advantage of this gauge over, for example, the synchronous gauge
is that the perturbed space curvature can be directly determined from the
freely decaying traceless stress.

Although Sasaki and Tanaka's
work does not include the fluid component, the form of the metrics
remains the same as the pure-scalar field case when the fluid 
is present.  One can show that the leading-order contribution 
from the fluid in this constant $T$ gauge 
is from the "00" component
and the trace of space components of the perturbed fluid
energy-momentum tensor.  Moreover, as
the perturbed $\delta\rho_f$ and $\delta p_f$ are related by the
underlying equation of state, $\delta\rho_f$ is simply
$c_0(\rho_f+p_f)$ with a constant $c_0$.  In addition, the
perturbed time-component of the fluid $4$-velocity $\delta U_0$
is simply $-\delta H/H$, the perturbed Hubble parameter;
the space component $\delta U_i$ appears only on the order of
$O(k/H)$ in the "0i" component of the perturbed energy-momentum
tensor in the same manner as the scalar fields do.   
Hence, inclusion of the fluid
retains all the good features of the constant-$T$ gauge
discussed by Sasaki and Tanaka for pure scalar fields. 
This is due partly to that the fluid dynamics can be formulated in 
a way very similar to that of the scalar field, as has been shown 
in the last section, and partly to that the formulation
using $T$ as a new time variable has already absorbed 
the gravity in the matter dynamics in a self-consistent manner.

Since the cosmological perturbations result from the departure of slightly 
different initial conditions, there exist $2(Nm+1)$ independent perturbations, 
due to the $2(Nm+1)$ independent 
phase-space coordinates for the $N$ scalar fields of 
$m$ components, plus a fluid component.  After one manages to make the
$2(Nm+1)$ scalar modes orthogonal, the space curvature of the constant-$T$ 
hypersurface only exists in two
canonically conjugated modes, and vanishes in all others.
These two modes are the adiabatic growing and decaying modes, and
the rest are entropy modes.
The growing adiabatic mode is associated with two trajectories
with exactly the same initial phase-space coordinates but differing by a 
$\Delta T_0$ in their initial launching times.  The two orbits
follow exactly the same phase-space trajectories, but at any given time 
$T$ their phase-space coordinates are different. 
This time-translational mode exists regardless of whether
the fluid is present.

Among all independent
perturbations, some may be integrable and thus stable, some are chaotic
and thus unstable, and the rest can still 
follow some stable manifolds in the 
phase space and thus be stable.  Note that the time-translation mode is
always stable.  For those perturbations that are integrable,
they are all associated with some integrals of motion $\alpha_j$'s, 
and result
from different trajectories with slightly different $\alpha_j$'s.
For example, the above Coulomb field gas has a conserved
total angular momentum, $L$, and there exists an integrable 
angular-momentum perturbation.  Since the Coulomb gas repels,
the "rotational motion" must slow down and hence the energy density of
the angular-momentum mode decays in time.
For those that follow stable orbital manifolds, these modes eventually
converge into some attractors since this dissipative nonlinear 
system has attractors in various regions of the $2Nm$ dimensional
phase space.   These modes are usually associated with the "velocity"
degree of freedom, as the presence of cosmic drag makes velocities 
eventually die out. 
By contrast, the chaotic unstable modes are usually associated with the 
"position" degree
of freedom since the initial transient can render the orbits
to end up at very different locations; these chaotic unstable modes 
give rise to perturbations in the potential energy $V_{eff}$. 
In great contrast to the single-field case, there are many
independent entropy modes in the multi-field
system, and many of these entropy modes can be more unstable 
than the adiabatic fluctuations.

As the multi-field dynamics is likely chaotic, the perturbations tend
to be entangled.  Though the constant-$T$ slice
provides a convenient way to calculate the cosmological
perturbations, the physically relevant
time slice is such that the perturbed energy-momentum tensor
has a vanishing $0i$ component, i.e., the comoving slice.
This is due to that the constant-$T$ gauge contains an extra gauge mode, 
which is always attached to the time-translational mode mentioned earlier.
This gauge mode can be removed in the comoving gauge.

The matter rest frame requires $T_0^i=0$, or
\begin{equation}
k[\sum_i\Delta\psi_i\psi_i'+\Delta\theta_f\theta_f'({n_f^2\over \rho_f+p_f})]=0,
\end{equation}
where $k$ is the small but finite wavenumber of the space Fourier mode.
It means that the difference of two
adjacent trajectories is constrained to be perpendicular to the
background "momentum" (c.f., Eq.(36)).
As $\Delta\psi_i$ and $\Delta\theta_f$ lie on a $Nm$ dimensional comoving 
hypersurface in the $Nm+1$-dimensional embedded space, this $Nm$
dimensional comoving time-slice is simply the
$S(\psi_i,\theta_f)|_T=const.$ hypersurface,
where $S$ is the Hamilton principal function.

We recall the discussions at the end
of last section that two adjacent orbits must be evaluated at different
times $\Delta T$ for them to lie on the same constant-$W_q$ hypersurface.
Now, the arguments should be extended to include the fluid, and $W_q$
is replaced by $S$.  
The difference of nearby trajectories $\delta\psi$ and $\delta\theta_f$ 
at the constant-$S$ surface therefore satisfy
\begin{equation}
\sum_i \psi_i'(\delta\psi_i(T)+\psi_i'\Delta T)+
({n_f^2\over\rho_f+p_f})\theta_f'(\delta\theta_f(T)+\theta_f'\Delta T)=0,
\end{equation}
thereby giving the time difference
\begin{equation}
\Delta T=-{\sum_i(\delta\psi_i\psi_i')+(n_f^2/\rho_f+p_f)\delta\theta_f\theta_f'
\over\sum_i(\psi_i')^2+(n_f^2/\rho_f+p_f)(\theta_f')^2}, 
\end{equation}
where $\delta\psi$ are $\delta\theta_f$ are the difference 
of two adjacent orbits evaluated
at the same $T$.  For the time translational mode, 
we have vanishing $\delta\phi_i$ and $\delta\theta_f$ when observed 
in the comoving slice, and hence
$\Delta T_0=\delta\psi_i/\psi_i'=\delta\theta_f/\theta_f'$, 
thereby yielding $\Delta T=-\Delta T_0$.
This cancels the original enhancement of curvature perturbation $R$ due to 
the difference of initial launching times $\Delta T_0$ in the
constant-T gauge, and reverts the curvature perturbation
back to its intrinsic value $R_0$ given by initial quantum fluctuations.    
That is, the contribution of $\Delta T_0$ to the perturbed
space curvature $R$ in this constant-T gauge 
is the gauge degree of freedom and removed 
in the comoving gauge\cite{10,16}.
 
\section{Conclusions and Discussions}

We formulate the coupled dynamics of quintessence, gravity and fluid 
in a unified framework by using the logarithmic cosmic expansion factor
$T(\equiv\ln a)$ as the new time variable.  The field dynamics turns out to
be identical to the dynamics of a relativistic particle immersed 
in a static viscous medium.  The presence of a fluid component serves 
merely to weaken the force acting on the field. 
Such formulation can be extended to describing many-field dynamics
as a many-body problem.  The forces on these fields can 
be made to be self-consistent ones 
by incorporating the mediate scalar bosons in the
field functional space.   This work also points out that when the number
of fields is large, it is possible to reduce the many-field problem to 
an effective single-field problem.  We work out the cold Coulomb field 
gas and warm Coulomb field gas as two examples.

It is interesting to note that when the multi-field dynamics is 
cast in the framework of relativistic Hamilton-Jacobi theory,
it can be straightforwardly
extended to quantum mechanics where Eq.(38) becomes
the Klein-Gordon equation with an evolving squared mass, $2gV_{eff}e^{6T}$ .  
This can be made exact when $\rho_f=0$
in the very early universe before photons were produced.  The quantization
are with respect to the scalar field $\psi_i$ relative to the conjugate
momentum $\pi_i$ (c.f., Eq.(36)), as well as the Hamiltonian $h$ relative
to the conjugated expansion factor $\ln a (=T)$
(c.f., Eq.(37)).  Note that the gravity has been incorporated in the
quantization since the action of Eq.(18) contains the gravity piece.  
Due to the presence a time-dependent squared mass in the
Klein-Gordon equation, it gives rise to low-energy ($\sim\hbar H$) 
quantum fluctuations of the quintessence
field and metrics on the superhorizon scale.  

The quantization of $\psi_i$ in the field
space logically fits well with our earlier treatment of 
the scalar bosons $V$, 
which mediate the quintessence interactions discussed in
Sec.(3), in the field functional space.  (Of course, quantization of the
scalar bosons $V$ can also be performed 
when necessary.)  As has been pointed out immediately below Eqs.(7) 
and (8), 
the ultra-relativistic regime renders the Hamiltonian $h$ to be a
good constant of motion.
In this regime, the interactions $V_{i,j}$ is comparatively small,
and the dominant energy component is the free field.
It thus provides
an easy way for quantization of $\psi_i$ together with the gravity
along the light cone.

The space coordinates have been dropped out of scene in the above analyses
because we are confined to the superhorizon scales.  It may be regarded
as a particular reference frame where the dynamics takes
place in the rest frame of the space coordinates (in contrast to the field 
coordinates) with a vanishing space momentum.  It remains to be investigated
whether the Planck-scale physics really admits such a symmetry that
one can generally transform the reference frame to the proper frame 
where the space momentum vanishes.  For example, when the uncertainty
principle
applies to the classical notions of space and time in the Planck era, 
space can become fuzzy in a reference frame where time has a definite
value.  If so, the Bose-Einstein condensation of a collection of 
$N$ such superhorion
quanta may constitute the foundation for the background $N$ quintessence
fields as well
as the gravity in the Planck scale, and provide the needed energy content
to develop into one universe.  

We now return to discussions on the classical quintessence fields.
Clearly, the dynamics of classical multi-field can easily 
be non-integrable and
chaotic.  Though a large number of fields help simplify
the problem, such a step is probably too drastic to take for a
possible extension beyond the single-field quintessence.  
It is probably more
natural to investigate dynamics of a few coupled fields together with
a fluid component.  However, this is surely a difficult 
regime as far as
the chaotic dynamics is concerned, since the coarse-grained averaged
quantities do not exist. Nevertheless, since the friction
applies uniformly in the velocity space, the acceleration may quickly
become negligible and the velocities become non-relativistic after an
initial transient, especially when the
fluid energy density is non-negligible and weakens the forces 
on the quintessence fields.  In such a "rolling" regime, the 
degrees of 
freedom are reduced by one half and the dynamics becomes much simpler.  
However, even after such reduction, if the remaining degrees
of freedom exceed or equal $3$, it is possible that there exists strange
attractors, where the trajectories stretch, twist and fold in a
multi-dimensional manifold, and one will again face a technically
insurmountable problem.  This can be so when the number of coupled
quintessence fields exceeds or equals $3$.

On the other hand, the scalar-field
dynamics in the early universe 
has gross imprints in the late-time cosmological
perturbations.  Some entropy perturbations
may have larger field amplitudes than the adiabatic perturbations do, 
if indeed multi-fields were active in the early universe.  
They generally yield a different fluctuation power spectrum in the 
cosmic microwave background radiation (CMBR) from the one predicted by
the pure adiabatic perturbation.  To be consistent with the observed
CMB temperature anisotropy spectrum, $k^n$ with $n\approx 1$ over three
orders of magnitude in length scale, or
in unit of the logarithmic scaling factor $\Delta T\sim 10$,
the entropy modes cannot be too unstable within this interval in the 
inflationary epoch.  
This is due, on one hand, to that the potential $V$ is flat during 
inflation and
the field dynamics is slow, and on the other hand to that the entropy modes
normally lead to a spectrum of 
density perturbations dominated by
by the large scale, but the COBE measurements have fixed their 
amplitudes to be less than that of the adiabatic perturbation\cite{17}.  
However,
the entropy mode may become unstable 
after the inflation.  Instabilities give rise to small-scale power
in the CMB fluctuations.  In view of the relatively large error bars in the CMB
measurements of fluctuation spectrum\cite{4,5,6}, 
there may still be rooms for accommodating
entropy modes resulting from the multi-field quintessence.


\begin{thebibliography}{99}
\bibitem{1} S. Perlmutter et al., Nature {\bf 391}, 51(1998); Astrophys. J.
{\bf 517}, 565(1999)

\bibitem{2} A.G. Riess et al., Astron. J. {\bf 116}, 1009(1998)

\bibitem{3} A.G. Riess et al., astro-ph/0104455; M.S. Turner \& A.G. Riess,
astro-ph/0106051; S. Jha, et al., astro-ph/00101521.

\bibitem{4} P. de Bernardis et al., Boomerang Coll., Nature {\bf 404},
955(2000) 

\bibitem{5} S. Hanany et al., Maxima Coll., Astrophys. J. {\bf 545}, L1-L4 (2000)

\bibitem{6} C. Pryke, N.W. Halverson, E.M. Leitch, J. Kovac, J.E. Carlstrom, W.L. Holzapfel \& M. Dragovan, astro-ph/0104490

\bibitem{7} R.R. Caldwell, R. Dave \& P.J. Steinhardt, Phys. Rev. Lett.
{\bf 80}, 1582 (1998); C. Wetterich, Nucl. Phys. B{\bf 302}, 668 (1988);
I. Zlatev, L. Wang \& P.J. Steinhardt, Phys. Rev. Lett. {\bf 82}, 986 (1999);
V. Faraoni, Phys. Rev. D{\bf 62}, 023504 (2000); W. Zimdahl, D.J. Schwarz, A.B. Balakin \& D. Pavon, Phys. Rev. D{\bf 64}, 063501 (2001); L. mendola, Phys. Rev. 
D{\bf 62}, 043511 (2000); L. Amendola \& D. Tocchini-Valentini, Phys. Rev. D{\bf 64},
043509 (2001); J.D. Barrow, R. Bean \& J. Magueijo, MNRAS {\bf 316}, L41 (2000) 

\bibitem{8} J.H. Traschen \& R.H. Brandenberger, Phys. Rev. D{\bf 42}, 2491 (1990)

\bibitem{9} L. Kofman, A.D. Linde \& A.A. Starobinsky, Phys. Rev. Lett. {\bf 73},
3195 (1994); Phys. Rev. D{\bf 56}, 3258 (1997)

\bibitem{10} M. Sasaki \& T. Tanaka, Prog. Theor. Phys. {\bf 99}, 763 (1998)

\bibitem{11} B. Ratra \& P.J.E. Peebles, Phys. Rev. D{\bf 37}, 3406 (1988)

\bibitem{12} T. Chiueh, Phys. Rev. E{\bf 49}, 1269 (1994) 

\bibitem{13} see, for example, J.D. Jackson, "Classical Electrodynamics", 
p. 580, Wiley, New York (1975), 2nd. Ed.
 
\bibitem{14} J.A. Gu \& W-Y.P. Hwang, Phys. Lett.{\bf B 517}, 1 (2001)

\bibitem{15} L.A. Boyle, R.R. Caldwell \& M. Kamionkowski, astro-ph/0105318

\bibitem{16} M, Sasaki \& E.D. Stewart, Prog. Theor. Phys. {\bf 96}, 
71 (1996); H. Komada \& T. Hamazaki, Phys. Rev. D{\bf 57}, 7177 (1998)

\bibitem{17} R. Strompor, K.M. Gorski \& A.J. Banday, 
Astrophys. J. {\bf 463}, 8 (1996).

\end{thebibliography}
\end{document}